\documentclass[12pt,a4paper]{article}
\usepackage{graphicx}
\usepackage{times}
\title{\bf Scientific collaborations in astronomy between amateurs and professionals}
\author{Johan H. Knapen$^{1,2}$\\
\\
\normalsize $^1$ Instituto de Astrof\'\i sica de Canarias, E-38200 La Laguna, Tenerife, Spain\\
\normalsize $^{2}$ Departamento de Astrof\'\i sica, Universidad de La Laguna, E-38205 La
Laguna, Tenerife, Spain
\\
\\
\normalsize Published in proceedings of \\
\normalsize"Stellar Winds in Interaction", editors T. Eversberg and J.H. Knapen. \\ 
\normalsize Full proceedings volume is available on http://www.stsci.de/pdf/arrabida.pdf
}
\date{\mbox{}}
\begin{document}
\maketitle
%
%
\def\bull{\vrule height .9ex width .8ex depth -.1ex}
\makeatletter
\def\ps@plain{\let\@mkboth\gobbletwo
\def\@oddhead{}\def\@oddfoot{\hfil\tiny\bull\quad
Workshop ``Stellar Winds in Interaction'' Convento da Arr\'abida, 2010 May 29 - June 2\quad\bull}%
\def\@evenhead{}\let\@evenfoot\@oddfoot}
\makeatother
%
%
\def\beginrefer{\section*{References}%
\begin{quotation}\mbox{}\par}
\def\refer#1\par{{\setlength{\parindent}{-\leftmargin}\indent#1\par}}
\def\endrefer{\end{quotation}}
%
%
{\noindent\small{\bf Abstract:} 
As our successful Mons campaign to observe WR140 has shown, there is a
strong interest among both amateur and professional astronomers to
collaborate on specific scientific questions. I highlight here some recent
examples of successful collaborations, and outline a number of areas of
astronomy where Pro-Am collaborations are making a difference. 
}
%
%
\section{Introduction}

Both professional and amateur astronomers have been studying the skies for centuries. Their respective r\^{o}les, however, have changed considerably. In the 18$^{\rm th}$ and 19$^{\rm th}$ century, for instance, one of the main tasks of professional astronomers was to calculate astronomical data to be used by the merchant fleet and the military, and to provide the society at large with important data such as the times of sunrise and sunset and, in fact, time itself. Amateur astronomers meanwhile entertained themselves with what most astronomers might now consider more interesting activities, such as discovering planets and comets, and observing nebulae. Well-known amateurs include Caroline Herschel (1750-1848) who discovered several comets, and her brother William (1738-1822) who discovered Uranus, several moons of that planet and of Saturn, created a catalogue of nebulae (a term used at the time to describe any extended object), and observed double stars. He constructed hundreds of telescopes, and made a number of important discoveries related to light and radiation. Needless to say, amateur astronomers at the time were wealthy individuals.

While in the 19$^{\rm th}$ century the interests of professional and amateur astronomers started to overlap more and more, they diverged again during the 20$^{\rm th}$ century. One of the main reasons for this is that professionals started to use expensive and exclusive telescopes, such as the ones used by Hale and Hubble in California, which were out of reach of all amateurs. 

We are now at the start of the 21$^{\rm st}$ century, and one of the characteristics of our modern times is the almost ubiquitous availability of high-quality yet relatively cheap technology. So whereas professional astronomers now use very advanced instruments, including very large optical and radio telescopes, and massive supercomputer power, interested amateurs can start to use technologically advanced telescopes, cameras, and computers at a cost which is accessible to many millions of citizens around the world. In the next Section of this paper, I will summarise some of the areas in which amateurs can, and do, use these tools to help advance professional astronomy. 

\section{Professional-Amateur Collaborations}

There are various areas of astronomy where amateurs can and do play a role. Almost without exception, these are based upon either the large amount of time that amateurs can dedicate, or the large amounts of independent observations or calculations they can collectively generate. I will give examples of some of these areas in the following subsections.

\subsection{Ultra-deep exposures}

In an era where professional astronomy keeps producing spectacular images of ``deep-space" objects (galaxies, planetary nebulae, etc.), it is perhaps surprising that amateur astronomers can still help by obtaining ultra-deep exposures. This is mainly due to the combination of two factors, namely, first, that the level of surface brightness reached in an image does not depend much on the aperture of a telescope, and, second, that professionals rarely can spend a night, or longer, on obtaining a deep image of one single target. When amateurs do precisely that, the results can be spectacular. 

An example of this is the ProAm collaboration between R. Jay Gabany and D. Mart\'\i nez-Delgado and their colleagues. They use the private, robotic, Blackbird Observatory in New Mexico with a 20\,inch RC Optics telescope to take deep images of nearby galaxies---reaching up to ten times fainter than images from the Sloan Digital Sky Survey (SDSS). This is achieved through a combination of a large field of view (a small telescope with a large CCD), long exposures, of typically 5-20\,hours, and the use of a luminance filter which allows the maximum amount of light to pass to the camera. This is done at a very dark observing site. 

\begin{figure}[ht]
\centering
\includegraphics[width=15cm]{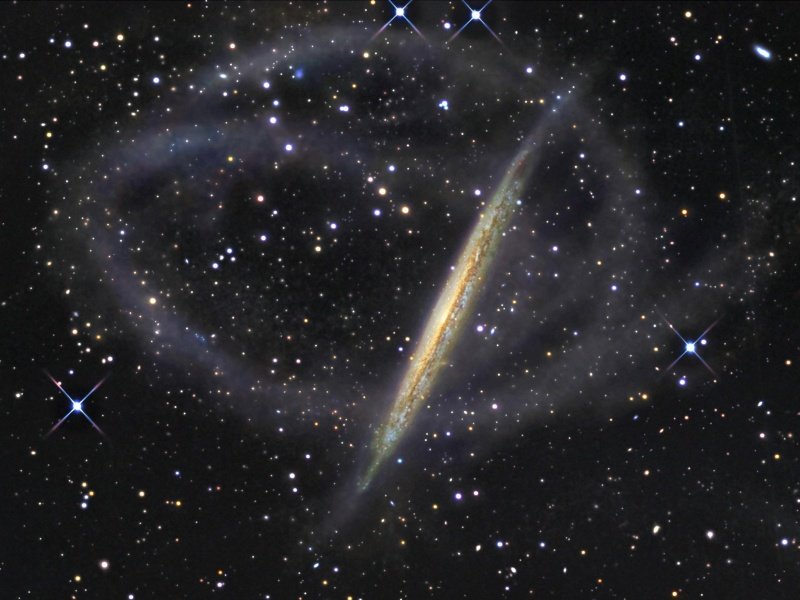}
\caption{Stellar streams outside the main disk of the edge-on galaxy NGC~5907, as seen on this deep optical image obtained at the amateur Blackbird Observatory. Reproduced from APOD (2008 June 19); image credit and copyright:  R. Jay Gabany (Blackbird Observatory)---collaboration; D. Mart\'\i nez-Delgado (IAC, MPIA), J. Pe\~{n}arrubia (U.Victoria), I. Trujillo (IAC), S. Majewski (U.Virginia), \& M. Pohlen (Cardiff).}
\end{figure}

As an example of their work, I reproduce in Fig.~1 their deep optical image of NGC~5907, a nearby galaxy with a set of conspicuous outer stellar streams. As described in detail in their paper (Mart\'\i nez-Delgado et al. 2008), these stellar streams are thought to originate in a past encounter of NGC~5907 with a smaller companion galaxy, and yield very interesting constraints on cosmological models of galaxy evolution. This team has subsequently published more results on such beautiful amateur images of nearby galaxies, and several of their images can be found on their website (www.cosmotography.com) or on the Astronomy Picture of the Day (APOD) site (apod.nasa.gov). They serve as an excellent illustration not just of what amateur astronomers with dedication and state-of-the-art equipment can achieve, but also of how a successful ProAm collaboration can lead to scientific progress.

\subsection{Meteors and Meteorites}

When a debris particle (a small or large rock called a {\it meteoroid}) enters the Earth atmosphere, it burns and leaves a light-emitting trail (a {\it meteor}) which can be easily observed from the ground. If the particle was large enough, part of it can survive its descend through the atmosphere and fall on the ground, in which case the remnant is called a {\it meteorite}. Meteorites are interesting because they can shine light on the chemical and physical conditions outside the Earth, and on the past history of the Solar System.

Observations of meteors and the recovery of meteorites depends on many different observers, mainly because the events are short-lived, not generally predictable, and very much localised geographically. These observers (and recoverers, to a lesser extent) cannot possibly all be professionals, and amateurs thus play an important role. Many different reports of a meteor can be collected and lead to the calculation of a rather detailed three-dimensional orbit trajectory. In case of a very bright meteor, or fireball, one can then deduce whether, and if so where, a meteorite may have fallen. A search operation can then be mounted to recover the meteorite(s).

A rather spectacular case was reported last year by Jenniskens et al. (2009). This team studied the impact of a small asteroid, called 2008 TC3, over northern Sudan on October 7, 2008. After collecting reports of the meteor sighting, the team calculated a trajectory and deduced the most likely place of impact of the resulting meteorite(s). A search party was then organised with the help of local residents and students, and a total of 47 meteorites with a total mass of 3.95\,kg were later recovered in this very sparsely populated desert region. Scientific results deduced from the meteorite recovery include the classification of the asteroid as F-class, most probably made up of a material so fragile that it had not previously been recovered from meteorites.

\subsection{Asteroid Shapes}

Even though they are small, asteroids moving in their orbits occasionally occult a star which is bright enough to be observed by amateur astronomers with modest optical equipment. Because many asteroid orbits are known to significant precision, as are the positions of bright stars, accurate predictions are available of when a certain asteroid will occult a star, and from where on Earth this should be observable. The observer sees the star reduce in brightness when it is occulted by the asteroid, for a certain duration of time depending directly on the linear size of the asteroid as it covers the sightline from the observer to the star.

\begin{figure}[ht]
\centering
\includegraphics[width=15cm]{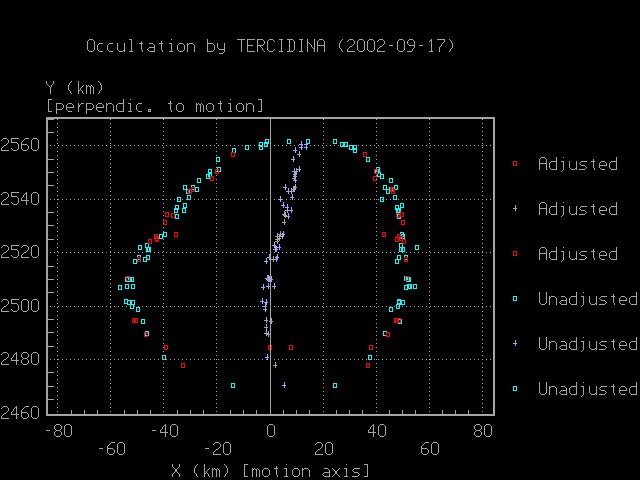}
\caption{Projected image of the asteroid Tercidina obtained from over 100 individual observations of the occultation by this asteroid of the star HIP~19388. From http://euraster.net/results/2002/20020917-Tercidina.html.}
\end{figure}

If large enough numbers of observers collaborate by observing the same occultation from a densely populated region of the world and then combining their data, an accurate map of the asteroid can be calculated. This is illustrated in Fig.~2, which shows a detailed image of the asteroid Tercidina, based on the collective observations of 105 European amateur observers in 17 countries, 75 of which measured the occultation (the remaining ``negatives" are equally useful to constrain the extent of the asteroid). The calculated image of Tercidina shows it to measure 92$\times$106\,km in size, and irregularities in its shape can be easily seen. Again, this is an example of many different amateur observers collaborating to reach a result that is impossible to obtain from one single viewpoint, even with the largest telescopes.

\subsection{Comets and Bright Supernovae}

The discovery of comets or bright supernovae has traditionally been a prime area for amateurs to contribute to astronomical scientific progress (also historically, e.g., Caroline Herschel). Nowadays, new asteroids and outer Solar system bodies are too faint to be discovered by amateurs, and are found through deep wide-area searches with professional observing programmes. But other ``new" objects, particularly comets and supernovae, are still waiting to be discovered, and some of those discoveries continue to be made by amateurs.

Before roughly 1996, of order ten new comets were discovered per year, and between 10 and 60\% of those were discovered by amateurs. New comets continue to be named after their discoverers, and some people become rather famous because of this. 

Nowadays, some 200 new comets are being discovered every year, around 1\% of which by amateurs (so in absolute numbers not much has changed: a few new amateur comets per year). The increase in numbers is thanks to new ways of observing. In particular, the {\it Solar and Heliospheric Observatory} ({\it SOHO}) satellite, which has been observing the Sun for over 15 years now, has discovered roughly half of all known comets, a total of up to 1900. This is because in its observations, it blocks out the direct light of the Sun, and as a by-product its images can reveal small comets passing close to the Sun. It is interesting that a fraction of {\it SOHO}'s comets have in fact been discovered by amateur astronomers, looking through the vast data sets produced by the satellite. Several amateurs have discovered more than 100 of these comets each...

With supernovae the story is similar. Once a matter of chance discovery by amateurs carefully observing the night sky, most supernovae are now found with automated, professional, observing experiments. This development has specifically been driven by cosmological studies, which use supernova statistics to constrain parameters such as the fraction of dark energy in the Universe, and thus our cosmological understanding of the Universe. Bright supernovae can still be discovered by amateurs by finding ``new" stars in galaxies or in certain regions of the sky. For instance, in 1987 the New Zealand amateur Albert Jones co-discovered the bright supernova 1987A in the large Magellanic cloud, which was the closest supernova for several hundred years.

\subsection{Time Series and Variability}

Although the introduction of queue scheduling helps to some extent, obtaining complete series of images or, especially, spectra of variable objects over long periods is becoming harder as telescopes get larger and access more competitive. One way out for some professionals is to build private dedicated telescopes, which are often robotic. But these are by their nature limited in the observing modes available, in access, and in the number and nature of objects that are observed.

Here is thus an interesting niche for enterprising amateur astronomers. They can nowadays make real contributions to scientific progress, mainly thanks to improvements in technology. This has lowered the cost and increased the availability of high-quality telescopes, spectrographs, CCD cameras, and data processing software. To make this efficient, amateurs should collaborate amongst themselves and with one or a few professionals who can provide scientific guidance, and lead the analysis and publication of the results obtained.

A very good example of such a collaboration is explained in detail in other contributions to these proceedings, which explain the various aspects of the Mons WR140 campaign. Such campaigns are technologically and intellectually challenging for those involved, and lead to scientific progress, as judged by contributions to the professional literature (e.g., Leadbeater \& Stencel 2010; Fahed et al. 2011; Morel et al. 2011).

\subsection{Classifying and Computing}

A relatively new way of amateurs directly contributing to progress in science (not just astronomy) has been facilitated by the widespread availability of computers in people's homes and broadband internet connectivity. This has led to two new kinds of endeavours, namely amateurs classifying huge numbers of scientific images, and large scientific computing projects being executed on individuals' home computers.

\subsubsection{Classifying by Amateur Volunteers}

Amateur classification in astronomy started off in 2000 with the Stardust@home project,  where volunteers could register to help NASA find impacts from interstellar dust particles on large series of images taken by the Stardust spacecraft. That was followed in 2007 by the extremely successful GalaxyZoo project (galaxyzoo.org; Lintott et al. 2008). The team of professional astronomers who developed this set up a website where volunteers could look at images of galaxies from the SDSS and answer a number of simple questions, such as whether the galaxy had discernible spiral structure or not. This was hugely successful, probably because contributing to the project could be done from the comfort of one's home, at any time of the day, because participating was simple and did not require any investment in terms of purchasing equipment, or training, and because not only did one contribute to ``real" science, there was also the chance to make significant discoveries. The fact that a volunteer might well be the first person ever to look in detail at a specific galaxy no doubt contributed to the success.

In the first year, GalaxyZoo collected some 50 million classifications from around 150000 volunteers around the world (many of these classifications were repeat observations, allowing the team to reach reliable results). By now, the team of professional astronomers behind GalaxyZoo has published around 20 scientific papers in the professional literature. 

Undoubtedly the most famous GalaxyZoo participant is a Dutch school teacher named Hanny van Arkel. In 
2007, she noticed a small blue-green patch of light below an SDSS galaxy and posted a message asking whether anyone knew what it was. No one knew, and it was called ``Hanny's Voorwerp" (Hanny's Object). She has since collaborated with a number of professional astronomers who have concluded from additional observations that the patch is most probably a cloud of gas illuminated by a quasar (J\'ozsa et al. 2009; Lintott et al. 2009; Rampadarath et al. 2010). Hanny's Voorwerp has made her famous, and her many media appearances make her an excellent ambassador of astronomy and of ``citizen science" in general.

The GalaxyZoo project has since evolved into the ``Zooniverse" (www.zooniverse.com), where thousands of volunteers continue to classify images of astronomical interest ranging from the surface of the Moon to supernovae and galaxy mergers.

\subsubsection{Distributed Computing}

Distributed computing uses a distributed system of computers which communicate through a network as a way of using spare computing capacity on many individual machines. In a number of areas of science, including astronomy, this is now being used very successfully to enlist the help of volunteers. The volunteer can contribute to a scientific computing project by registering as a participant on the web, and then allowing the project to execute calculations on his or her home computer. This is done when the computer has spare capacity, and often people leave their computers on at night. Volunteers thus contribute a bit of their time, and some electricity costs. 

The area started in 1996 with prime number searches. In the field of astronomy, we can consider the SETI@home (setiathome.berkeley.edu; SETI stands for Search for Extra-Terrestrial Intelligence) project to be the pioneer. Starting in 1999, it has been using massively distributed computing to analyse large amounts of radio data on volunteers' home computers. To date, no signatures from life in space have been detected, but SETI@home is listed in the Guinness book of records as the largest computation in history. It has over 5 million participants world-wide, who no doubt all hope to own the computer that will process the first sign of extraterrestrial life. 

The PlanetQuest programme (www.planetquest.org) tries to find extrasolar planets by reducing stellar imaging from major telescopes on home computers made available by volunteers. In other areas of science, the Folding@home (folding.stanford.edu) project is noteworthy: it uses massive amounts of computing power provided by volunteers to study protein folding and molecular dynamics.

\section{Concluding Remarks}

In astronomy, as in other areas of science, ``citizen science" is important because it allows dedicated amateurs to contribute directly to scientific progress in collaboration with professionals, but also because it is a great way to popularise science. Many amateurs collaborate from the comforts of their home through computer-based projects such as the Zooniverse or SETI@home. Others are more traditional observers, some of whom have teamed up to provide valuable meteor, comet, or asteroid observations. Yet others, including the authors of other papers in these proceedings, use spectroscopic monitoring of interacting stars to provide professionals with data which are hard to obtain with professional telescopes. 

Joining existing ProAm collaborations is sometimes possible for interested amateurs. Starting new ones depends on meeting the right professional astronomer, and may not be easy because there must be a good match between the quantity and quality of the data that the amateur can offer and what the professional needs to advance his or her scientific research. In general, such collaborations can be very fruitful, as the examples described above clearly illustrate.

%

\footnotesize
\beginrefer

\refer Fahed, R., et al. 2011, in Proc. IAUS272 Active OB Stars: Structure, Evolution, Mass Loss and Critical Limits, in press

\refer Jenniskens, P., et al.\ 2009, Nature, 458, 485

\refer J{\'o}zsa, G.~I.~G., et al.\ 2009, A\&A, 500, L33

\refer Leadbeater, R., \& Stencel, R. 2010, arXiv:1003.3617

\refer Lintott, C.~J., et al.\ 2008, MNRAS, 389, 1179 

\refer Lintott, C.~J., et al.\ 2009, MNRAS, 399, 129

\refer Mart{\'{\i}}nez-Delgado, D., Pe{\~n}arrubia, 
J., Gabany, R.~J., Trujillo, I., Majewski, S.~R., 
\& Pohlen, M.\ 2008, ApJ, 689, 184

\refer Morel, T., et al. 2011, in Proc. IAUS272 Active OB Stars: Structure, Evolution, Mass Loss and Critical Limits, in press 

\refer Rampadarath, H., et al.\ 2010, A\&A, 517, L8 

\endrefer           

\end{document}